\documentclass{llncs}

\usepackage[T1]{fontenc}
\usepackage{amsmath}
\usepackage{amsfonts}
\usepackage{mathtools}
\usepackage{stmaryrd}
\usepackage[bb=boondox]{mathalfa}

\usepackage{graphicx}
\usepackage{multicol}
\usepackage{ifdraft}
\usepackage{ifthen}
\usepackage{ebproof}
\usepackage{algorithm}
\usepackage{algorithmicx}
\usepackage{xspace}
\usepackage[noend]{algpseudocode}
\usepackage{tikz, pgf}
\usepackage{url}
\usepackage{subcaption}
\usepackage[short]{lib/latexutils/abbrev}
\usepackage{lib/latexutils/sidenote}
\usepackage{lib/latexutils/symbols}
\usepackage{lib/latexutils/theorems} 
\usepackage{booktabs}
\usepackage{tabularx}

\usepackage{hyperref}

\usepackage[subtle,
    bibbreaks=tight,
    mathspacing=tight,
    floats=tight,
    wordspacing=tight,
]{savetrees}

\usepackage[frozencache,cachedir=_minted]{local/minted}

\newcommand{\python}[1]{\mintinline{python}{#1}}

\setminted{
  linenos=true,
  autogobble
}

\ebproofset{
  right label template=\textsc{\inserttext}
}
\ebproofnewstyle{small}{
  separation = 1em, rule margin = .5ex,
  template = \footnotesize$\inserttext$,
  right label template=\footnotesize\textsc{\inserttext}
}

\usetikzlibrary{cd, arrows.meta, decorations.pathmorphing}
\tikzset{
  >=Kite,
  invisible/.style={opacity=0},
  visible on/.style={alt={#1{}{invisible}}},
  alt/.code args={<#1>#2#3}{%
      \alt<#1>{\pgfkeysalso{#2}}{\pgfkeysalso{#3}}%
  }
}
\tikzcdset{
  arrow style=tikz,
  diagrams={>=Kite},
  ampersand replacement=\&,
  labels=description
}

\algrenewcommand\algorithmicindent{1.0em}%
\algrenewcommand\algorithmicfunction{\textbf{fn}}
\algnewcommand\algorithmicyield{\textbf{yield}\ }
\algnewcommand\algorithmicyieldfrom{\textbf{yield}\ \textbf{from}\ }
\algnewcommand\algorithmiccontinue{\textbf{continue}}
\algnewcommand\algorithmicbreak{\textbf{break}}

\newcommand\Continue{\algorithmiccontinue}

\newcommand\algcomment[1]{\hfill{\textcolor{gray}{(#1)}}}

\algdef{SE}[DOWHILE]{Do}{DoWhile}{\algorithmicdo}[1]{\algorithmicwhile\ #1}

\DeclareMathOperator*{\powerset}{\ensuremath{\mathcal{P}}}
\newcommand{\tstr}{\ensuremath{\mathtt{str}}}

\ifthenelse{\boolean{true}}{%

}{%

}

\newboolean{preprint}
\newboolean{submission}

\renewcommand\eps{\ensuremath{[]}}

\newcolumntype{C}{>{\centering\arraybackslash}X}

\setboolean{preprint}{true}
\setboolean{submission}{false}

\begin{document}
  \title{Optimized Execution of FreeCHR}
\author{Sascha Rechenberger \and Thom Frühwirth}
\authorrunning{S. Rechenberger, T. Frühwirth}
\institute{%
    Institute of Software Engineering and Programming Languages,\\
    Ulm University, Albert-Einstein-Allee 11, 89069 Ulm, Germany\\
    \email{\{sascha.rechenberger,thom.frühwirth\}@uni-ulm.de}
}

  \maketitle
  \begin{abstract}
  \emph{Constraint Handling Rules} (CHR) is a rule-based programming language that rewrites collections of constraints.
  It is typically embedded into a general-purpose language.
  There exists a plethora of implementations for numerous host languages.
  However, the existing implementations often re-invent the method of embedding, which impedes maintenance and weakens assertions of correctness.
  To formalize and thereby standardize the embedding of a ground subset of CHR into arbitrary host languages, we introduced the framework \emph{FreeCHR} and proved it to be a valid representation of classical CHR.
  For the sake of simplicity, abstract implementations of our framework did not yet include a concrete matching algorithm nor optimizations.
  In this paper, we introduce an improved execution and matching algorithm for FreeCHR.
  We also provide empirical evaluation of the algorithm.
  \ifthenelse{\boolean{preprint}}{This is a preprint of a paper submitted to the \emph{39th Workshop on (Constraint and Functional) Logic Programming (WLP 2025)}.}{}
\end{abstract}


  \section{Introduction}\label{sec::freechr:introduction}
\emph{Constraint Handling Rules} (CHR) is a rule-based programming language which is typically embedded into a general-purpose language.
Having a CHR implementation available enables software developers to solve problems in a declarative and elegant manner.
Aside from the obvious task of implementing constraint solvers \cite{fruehwirth2006complete,dekoninck2006inclp}, it has been used, \EG to solve scheduling problems \cite{abdennadher2000university} and implement concurrent and multi-agent systems \cite{thielscher2002reasoning,thielscher2005flux,lam2007concurrent}.
In general, CHR is ideally suited for any problem that involves the transformation of collections of data.
Programs consist of a sequence of rewriting rules which hide the process of finding matching values to which rules can be applied.
As a result, developers are able to implement algorithms and applications in a purely declarative way, without the otherwise necessary boilerplate code.

The literature on CHR as a formalism consists of a rich body of theoretical work, including a rigorous formalization of its declarative and operational semantics \cite{sneyers2010time,fruehwirth2015constraint}, relations to other rule-based formalisms \cite{fruehwirth2025principles} and results on properties like confluence \cite{abdennadher1996confluence,christiansen2015confluence,gall2017decidable}.
Implementations of CHR exist for a number of languages, such as Prolog \cite{schrijvers2004leuven}, C \cite{wuille2007cchr}, C++ \cite{barichard2024chr}, Haskell \cite{chin2008typesafe,lam2007concurrent}, JavaScript \cite{nogatz2018chr} and Java \cite{abdennadher2002jack,vanweert2005leuven,ivanovic2013implementing,wibiral2022javachr}.

We intend to formalize and thereby standardize the way of implementing CHR as an internal DSL for arbitrary host languages.
In our view, this will have three major advantages.
First, by formalizing CHR via \emph{initial algebra semantics}, we provide a tighter connection between the formal definition and the implementations.
Initial algebra semantics are a common concept in functional programming and are used to inductively define and implement languages and their semantics \cite{hudak1998modular,johann2007initial}.
This provides a common formal foundation for considerations concerning theory and implementations.
Secondly, by embedding CHR as an internal language, distribution and use are simplified.
Unlike embeddings which rely on an external source-to-source compiler, internal embeddings can be easily distributed and imported via the host languages package and module system.
They also do not add another link to the build chain which is, even with modern build tools, an additional nuisance. 
Finally, by providing a formal, host language agnostic definition of a CHR implementation, we also provide a generalized documentation for implementations.
We hope that this will ease contribution to future implementations and help to keep them maintained.

\emph{FreeCHR} \cite{rechenberger2025freechr} formalizes the embedding of a ground subset of CHR by using \emph{initial algebra semantics}.
It provides both, a guideline and high-level architecture to implement and maintain CHR implementations across host languages, and a strong connection between the practical and formal aspects of CHR.
We recently introduced an execution algorithm which implements the \emph{refined} operational semantics for FreeCHR and serves as an algorithmic representation of it \cite{rechenberger2025instance}.
However, for the sake of simplicity, we did not yet include a concrete matching algorithm nor optimizations.

In this work, we want to introduce and evaluate an improved execution algorithm which implements two optimizations, namely \emph{iterator based matching}, as described by Van Weert \cite{vanweert2010efficient}, as well as \emph{indexing}.
The idea behind iterator based matching is to generate a (lazy) sequence of all possible matchings once, instead of searching for a single matching every time.
The idea behind indexing is to associate sets of values with quickly (typically, in amortized constant time) accessible indices which are used to narrow the search space for possible matching values.
The matching algorithm then only needs to search the indexed set instead of all values.
To evaluate our optimizations, we implemented our algorithm in Python and compared its performance to the \emph{CHR} implementation provided by \emph{SWI-Prolog}.

The rest of the paper is structured as follows. 
\autoref{sec:preliminaries} introduces the syntax of FreeCHR programs.
\autoref{sec:exec} introduces an improved execution algorithm.
\autoref{sec:eval} describes experiments to evaluate the impact of the optimizations and discusses results.
Finally, \autoref{sec:conclusion} concludes the paper and discusses future work.

  \section{FreeCHR}\label{sec:preliminaries}
FreeCHR is a framework to standardize the embedding of ground CHR into arbitrary programming languages.
The main idea is to provide an inductive definition of CHR programs as algebraic expressions of the host language.
The abstract syntax of FreeCHR is defined as follows:
a single rule
  $\chrrule(n, \left[k_1,...,k_n\right], \left[r_1,...,r_m\right], g, b)$
is a FreeCHR program, where the \emph{name} of the rule $n \in \tstr$ is a string,
the \emph{kept head} $\left[k_1,...,k_n\right]$\footnote{%
We use Haskell-like syntax to denote lists: $[]$ is the empty list and $\mathit{x}:\mathit{xs}$ constructs a list with head element $\mathit{x}$ and tail $\mathit{xs}$. We will use the notations $\left(a : b : c : []\right)$ and $\left[a, b, c\right]$ interchangeably as we consider it useful.}
and \emph{removed head} $\left[r_1,...,r_m\right]$ are sequences of functions, mapping elements of $C$ to booleans.
$C$ is the domain, over which a FreeCHR program is defined.
We will call the functions $k_1$ to $r_m$ \emph{patterns} and use boolean functions to model them.
Since FreeCHR programs are defined over arbitrary host-language values, it is impossible to defined them as Herbrand-terms, like in classical CHR.
The \emph{guard} $g$ is a function, mapping sequences of elements of $C$ to booleans
and the \emph{body} $b$ is a function, mapping sequences of elements of $C$ to a new sequence of elements of $C$.
If $p_1$ to $p_l$ are FreeCHR programs, then the \emph{composite}
  $p_1 \chrcomp ... \chrcomp p_l$
is also a FreeCHR program.

\begin{example}[Euclidean algorithm]\label{ex:freechr:gcd}
    The CHR program
    \begin{align*}
      \mathtt{zero}\ &@\ 0\ \Longleftrightarrow \btrue\\
      \mathtt{subtract}\ &@\ N\ \setminus\ M\ \Longleftrightarrow\ \left. 0 < N, N \leq M\ \mid\ M-N\right. 
    \end{align*}
    which computes the greatest common divisor of a collection of numbers, can be expressed in FreeCHR by the program $\mathit{gcd} = \mathit{zero} \chrcomp \mathit{subtract}$ with
    \begin{align*}
        \mathit{zero} &= rule(\mathtt{zero}, \eps, \left[\lambda n. n = 0\right], (\lambda n. \btrue), (\lambda n. \eps)) \\
        \mathit{subtract} &= rule(\mathtt{subtract}, \left[\lambda n. 0 < n\right], \left[\lambda m. 0 < m\right],
            (\lambda n\ m. n \leq m), (\lambda n\ m. \left[m-n\right]))
    \end{align*}
    $\lambda$-abstractions are used for ad-hoc definitions of functions.
    To reduce formal clutter, we write the functions of guard and body as $n$-ary functions instead of unary functions, \IE $(\lambda n\ m. n \leq m)$ instead of $(\lambda \left[n,m\right]. n \leq m)$.
    
    When executed, the \textit{zero} rule will remove any elements $n = 0$.
    The \textit{subtract} rule replaces any elements $0 < m$ for which there is another element $0 < n$ and $n \leq m$ by the difference $m-n$.
\end{example}

An instance of FreeCHR is implemented by providing concrete implementations for the abstract symbols $\chrrule$ and $\chrcomp$.

  \section{Improved Execution Algorithm for FreeCHR}\label{sec:exec}
We will now introduce an improved FreeCHR execution algorithm.

\subsection{States}
We first model the structure of the states required for implementing the execution algorithm.
A state is a quadruple
\begin{align*}
    \langle Q, S, I, M\rangle \in \flist ((\tnat \setprod \mathtt{iter} \cup \tone) \setprod C)
        \ \setprod\  \powerset (\tnat \setprod C)
        \ \setprod\  \tnat
        \ \setprod\  \powerset{\left(I \setprod \tnat\right)}
\end{align*}
with $\mathtt{iter} = \tnat \setprod \flist (\flist (\tnat))$.
$\powerset\left(X\right)$ denotes the powerset of a set $x$, $\tone$ is a singleton set $\left\{\bot\right\}$, $\tnat$ are the natural numbers and $C$ is the domain of values, FreeCHR programs are defined over.
We call $Q$ the \emph{query}, $S$ the \emph{store}, $I$ the \emph{index}, and $M$ the \emph{index relation}.
Furthermore, we define the following functions to extract the elements of a state.
We denote the set of all states over a domain $C$ as $\Omega_I C$.
\begin{align*}
    \mathtt{query}_C\langle Q, \_, \_, \_\rangle &= Q & \mathtt{store}_C\langle \_, S, \_, \_ \rangle &= S\\
    \mathtt{index}_C\langle \_, \_, I, \_ \rangle &= I & \mathtt{index\_rel}_C\langle \_, \_, \_, M \rangle &= M
\end{align*}

The \emph{query} drives execution of programs.
It is modelled as a stack of decorated values.
The value on top of the query is called the \emph{active} value.
If a value on the query was the active value once, it is called \emph{activated}.
Values on the query are decorated by either $\bot$ if they were never active,
or by their unique identifier and an iterator.
An iterator consists of the index of the currently considered rule and a (lazy) sequence of possible configurations.
The functions 
\begin{align*}
    \mathtt{push\_query}_C&\left(\langle Q, S, I, M\rangle, c_1, ..., c_n\right) = \langle \left(\bot, c_1\right):...:\left(\bot, c_n\right):Q, S, I, M \rangle\\
    \mathtt{pop\_query}_C&\langle (i, c):Q, S, I, M\rangle = \langle Q, S, I, M\rangle
\end{align*}
define basic stack operations on the query of a state.
Note, that $\mathtt{pop\_query}_C$ is a partial function, defined only if the query is not empty.
It is up to the implementor, to handle the undefined case (\IE throw an exception or stay with the default handling provided by the programming language).
The functions
\begin{align*}
    \mathtt{get\_active\_iterator}_C&\langle (\_, \mathit{it}, \_):Q, S, I, M\rangle = \mathit{it}\\
    \mathtt{set\_active\_iterator}_C&(\langle (i_a, \_, c_a):Q, S, I, M\rangle, \mathit{it}) = \langle (i_a, \mathit{it}, c_a):Q, S, I, M\rangle
\end{align*}
define additional operations to access and modify the iterator of the active value.

The \emph{store} is a set of pairs of unique identifiers and values.
Upon activation, a value is uniquely identified and added to the store.
In constraint logic programming, the store is viewed as the collection of currently known facts, on which the constraint solver operates.
The function
\begin{align*}
    \mathtt{activate}_C&\left(\langle (\bot, c):Q, S, I, M\rangle, c\right)
        = \langle ((I, 1, []), c):Q, S \cup \left\{(I, c)\right\}, I+1, M' \rangle\\
    \text{where}\ M' &= \left\{
        \begin{matrix*}[l]
            \left\{\mathit{index}(c) \mapsto I\right\} \cup M &\quad& \text{if \textit{index} is defined on $c$} \\
            M &\quad& otherwise
        \end{matrix*}
    \right.
\end{align*}
adds the top value of the \emph{query} to the \emph{store}, with the current \emph{index} as the unique identifier and increments it by $1$ if this value was not activated before.
It also replaces the value with a decorated version and associates the identifier of the value with an index computed with the function $\mathit{index}$, if it is defined on $c$.
The \emph{index relation} is a relation of unique identifiers and indices.
If there is an indexed pattern requiring an index $\mathit{ix}$, the index map allows to reduce the search space for patterns to the set of values identified by indices $i$ with $\mathit{ix} \mapsto i$.
The function
\begin{align*}
    \mathtt{remove}_C&\left(\langle Q, S \cup \left\{(i_1, c_1), ..., (i_n, c_n)\right\}, I, M\rangle, i_1, ..., i_n\right) = \langle Q, S, I, M\rangle\\
    \text{where}\ M' &= M \setminus \left\{ix \mapsto i \in M \mid i \in \left\{i_1, ..., i_n\right\}\right\}
\end{align*}
removes the values with the given identifiers from the store and cleans up the index relation.
Finally, the function
\begin{align*}
    \mathtt{alive}_C&\left(\langle Q, S, I, M\rangle, i\right) = \left\{\begin{matrix*}[l]
        \btrue &\quad  & \exists c \in C. (i, c) \in S\\
        \bfalse &\quad  & \text{otherwise}
    \end{matrix*}
    \right.
\end{align*}
is used to check if a value with a certain identifier is an element of the \emph{store}.

\subsection{Applied Optimizations}
\emph{Iterator based matching} is mainly inspired by Van Weert \cite{vanweert2010efficient}.
The idea is to compute a \emph{lazy sequence of all} applicable matchings upon activation of a value.
Every time the value becomes active (including the first time it was activated) one applicable matching is read from the iterator.
When the iterator is empty, we know that there are no more applicable configurations.
Furthermore, by ensuring that we only take matchings which are applicable at the time of activation, \IE, only pairing the active value with ``older'' values, we automatically prevent reapplication \cite{vanweert2010efficient}.

In our implementation, we instantiate an iterator for each rule and save the index of the rule together with the iterator.
If all matchings for a rule are consumed and the value not removed, the index is incremented.


\begin{example}
Given a state $\langle [\left((2, 0, []), 12\right)], \left\{(0, 6), (1, 9)\right\}, 3, \emptyset \rangle$ and the program from \autoref{ex:freechr:gcd}.
Upon initializing the active value for the rule \emph{subtract}, the iterator will be set to $(1, \left[(0, 2), (1, 2)\right])$ or $(1, \left[(1, 2), (0, 2)\right])$, depending on how the store is traversed. \IE, we transition to either $\langle [\left((2, 1, \left[(0, 2), (1, 2)\right]), 12\right)], \left\{(0, 6), (1, 9)\right\}, 3, \emptyset \rangle$ or $\langle [\left((2, 1, \left[(1, 2), (0, 2)\right]), 12\right)], \left\{(0, 6), (1, 9)\right\}, 3, \emptyset \rangle$.
\end{example}

\emph{Indexing} is a technique that associates a value with a quickly accessible index to narrow the search space for matchings.
Since each pattern of a FreeCHR rule is an isolated arbitrary host-language function, it is in general\footnote{Exceptions may be highly symbolic languages like LISP or Prolog.} not trivial to automatically infer if there might be indexed values for a pattern.
We introduce a notation to manually decorate a pattern to declare how the index of matching values can be computed from another matched value.

For a given domain and program, the programmer needs to do two things.
First, they need to declare how a value of a certain type can be indexed by defining the function $\mathit{index}$ for those values.
Secondly, they need to add decorations to the patterns of their program.

Formally, a \emph{decorated pattern} is a triple $\left(\langle i_p; f_p \rangle @ h\right) \in \tnat \setprod \mathit{I}^{C} \setprod \tbool^{C}$, where $\mathit{I}$ is the set of indices used in the index relation.
We will assume that applying a decorated pattern to a value applies the function $h$, \IE, $\left(\langle i_p ; f_p \rangle @ h\right)(v) \equiv h(v)$.
The natural number $i_p$ declares the position of the referenced value in a (partial) matching, the function $f_p$ declares how the index to lookup is computed from the referenced matched value.
$h$ is the pattern function.

\begin{example}\label{ex:indexing}
    Given the FreeCHR rule
    \begin{align*}
        \textit{rule}(
            &\mathtt{transitive\_paths}, \left[\langle 1; \lambda \mathtt{path}(s, \_, \_, \_). s\rangle @ (\lambda \_. \btrue),\ \lambda \_. \btrue\right],\ \left[\right],\\
            &\lambda\ \mathtt{edge}(\_, y, \_)\ \mathtt{path}(y', \_, \_, \_).\ y = y',\\
            &\lambda\ \mathtt{edge}(x, \_, w_e)\ \mathtt{path}(\_, z, w_p, ns).\ \left[\mathtt{path}(x, z, w_e+w_p, x:ns)\right]
        )
    \end{align*}
    which is part of the algorithm to compute the shortest paths in a graph.
    For an edge $\mathtt{edge}(x, y, w_e)$ and a path $\mathtt{path}(y, z, w_p, ns)$, it adds the transitive path $\mathtt{path}(x, z, w_e + w_p, x:ns)$, with accumulated weight and an extended list of nodes.
    The decorated pattern $\langle 1; \lambda \mathtt{path}(s, \_, \_, \_). s \rangle @ (\lambda \_. \btrue)$ declares, that the index for possible values matching this pattern is computed by taking the source of the $\mathtt{path}$ matched to the pattern in the $1^{\text{st}}$ position of the head (beginning with 0).
    The programmer needs to define a function $\mathit{index} (\mathtt{edge}(\_, t, \_)) = t$.
    In practice, this can be achieved by, \EG, implementing a certain function, interface or super-class.

    Given a state with store $\left\{(0, \mathtt{edge}(a, b, w_0)), (1, \mathtt{edge}(b, c, w_1)), (2, \mathtt{path}(b, c, w_2, [b, c]))\right\}$ and an index relation $\left\{b \mapsto 0, c \mapsto 1\right\}$.
    Assuming, $\mathtt{path}(b, c, w_2, [b, c])$ is the active value, we quickly find that the only relevant value in the store is $(0, \mathtt{edge}(a, b, w_0))$, by applying the provided function to the active value.
\end{example}

\subsection{Implementation}
We now introduce the improved execution algorithm, starting with matching.

\begin{algorithm*}[t]
    \begin{algorithmic}[1]
        \Function{match}{$(h_1, ..., h_n)$,$\mathit{guard}$,$i_a$,$c_a$,$\mathit{state}$}
            \State $\mathtt{valid\_indices} \leftarrow \left\{i \mid \left(i, \_\right) \in \mathtt{store}\left(\mathit{state}\right)\right\}$
                \algcomment{Initialization}
                \label{alg:match:init:begin}

            \State $\left[I_1, ..., I_{n}\right] \leftarrow \mathtt{array}\left(n\right)$; $\left[V_1, ..., V_{n}\right] \leftarrow \mathtt{array}\left(n\right)$ 
                \label{alg:match:init:end}
            \State
            \Function{match\_rec}{$i$, $j$}
                    \label{alg:match_rec:begin}
                \If{$i = j$}\algcomment{Case 1: active value}
                        \label{alg:match_rec:case_1:begin}
                    \If{$h_j = (\langle r, f \rangle @ \_) \wedge r \geq j \wedge f\left(V_r\right) \not\mapsto i_a$}
                            \State \Return $[]$
                    \EndIf
                    \State $I_j \leftarrow i_a$; $V_j \leftarrow c_a$
                    \State \Return $\textsc{match\_rec}(i, j-1)$
                        \label{alg:match_rec:case_1:end}
                \ElsIf{$j < 1$}\algcomment{Case 2: matching complete}
                    \label{alg:match_rec:case_2:begin}
                    \State \Return \textbf{if} $\mathit{guard}(V_1, ..., V_{n})$ \textbf{then} $\left[\left(i, I_1, ..., I_{n}, V_1, ..., V_{n}\right)\right]$ \textbf{else} $[]$
                    \label{alg:match_rec:case_2:end}
                \Else\algcomment{Case 3: split}
                    \label{alg:match_rec:case_3:begin}
                    \State $\mathit{indices} \leftarrow \mathtt{valid\_indices}$
                    \If{$h_j = (\langle r, f \rangle p) \wedge r \geq j$}
                        \State $\mathit{indices} \leftarrow \left\{\, i \mid f\left(\mathit{V}_{r}\right) \mapsto i\, \right\}$
                            \algcomment{overwrite initialization}
                    \EndIf
                    \State $\mathit{matchings} \leftarrow []$
                    \For{$i_s \in \mathit{indices}$}
                        \If{$i_s \geq i_a \bor i_s \in \left\{I_{j+1}, ..., I_{n}\right\}$}
                            \Continue
                        \EndIf
                        \State $c_s \leftarrow \mathit{lookup}(\mathtt{store}(\mathit{state}), i_s)$ 
                        \If{$\neg h_j(c_s)$}
                            \Continue
                        \EndIf

                        \State $I_j \leftarrow i_s$; $V_j \leftarrow c_s$
                        \State $\mathit{matchings} \leftarrow \mathit{matchings} \diamond \textsc{match\_rec}(i, j-1)$
                    \EndFor
                    \State \Return $\mathit{matchings}$
                    \label{alg:match_rec:case_3:end}
                \EndIf
            \EndFunction
                \label{alg:match_rec:end}

            \State
            \State \Return $\left[m \mid i \in \left[n, ..., 1\right], h_i\left(c_a\right), m \in \textsc{match\_rec}(i, n)\right]$
                \label{alg:match:end}
        \EndFunction
    \end{algorithmic}
    \caption{Refined matching algorithm}
    \label{alg:match}
\end{algorithm*}

The function \textsc{match} gets the head and guard of the rule, its currently active value $c_a$ and its index $i_a$, as well as the current state $\mathit{state}$ passed.
Lines \ref{alg:match:init:begin} to \ref{alg:match:init:end} initialize necessary values for the computation.
$\mathtt{valid\_indices}$ is the set of indices of all values which are currently saved in the store.
$[I_1, ..., I_n]$ and $[V_1, ..., V_n]$ are arrays of length $n$.
They are used to store the current (partial) matching during computation.
The function \textsc{match\_rec} (lines \ref{alg:match_rec:begin} to \ref{alg:match_rec:end}) performs the backtracking search.
The final line \ref{alg:match:end} returns the sequence of all matchings found by matching the active value $(i_a, i_c)$ to each of the head patterns from right to left.

The function \textsc{match\_rec} accepts the index for the active value $i$, as well as the running index $j$.
The function handles three cases.
The first case (lines \ref{alg:match_rec:case_1:begin} to \ref{alg:match_rec:case_1:end}) occurs if the designated position for the active constraint is reached.
First it is checked if the respective pattern is a \emph{decorated pattern} and if the reference index is not to the left of the current index.
If $h_j$ is an decorated pattern, it is necessary for $i_a$ to be in relation with the index $\mathit{ix}$ which is computed by applying the indexing function $f$ to the value matched to the head pattern at the annotated reference position $r$.\footnote{We will write $x \mapsto i$ instead of $x \mapsto i \in \mathtt{index\_rel}(\mathit{state})$ and $x \not \mapsto i$ instead of $x \mapsto i \notin \mathtt{index\_rel}(\mathit{state})$ respectively.}
If this is not the case, there is no valid matching with $c_a$ matched to $h_j$ and the current value of $V_r$ matched to $h_r$.
We then return an empty sequence.
Otherwise, we set $I_j$ to $i_a$ and $V_j$ to $c_a$, and continue the search with the next pattern.
In the second case (lines \ref{alg:match_rec:case_2:begin} to \ref{alg:match_rec:case_2:end}), we have collected a complete matching and check if it satisfies the guard.
If this is the case, we return a singleton sequence containing the matching, otherwise an empty sequence.
Finally, in case 3 (lines \ref{alg:match_rec:case_3:begin} to \ref{alg:match_rec:case_3:end}), we consider all candidates for the current pattern $h_j$.
We first initialize all currently alive values as possible candidates.
If $h_j$ is an decorated pattern, we can constrain the search space to values related to the computed index $f\left(\mathit{V}_{r}\right)$.
Otherwise, we loop over all candidates.
We first check if the candidate was activated \emph{before} the active value (to prevent reapplication) and if it was not used in the current partial matching before.
Then we look up the value $c_s$ associated with $i_s$ and check if it satisfies the pattern $h_j$.
If all this is true, we save identifier and value and descend further into the recursion.
The returned results are added to the sequence of all matchings found by this loop until now.\footnote{The operator $\diamond$ denotes the concatenation of two sequences.}

Note, that the algorithm presented in \autoref{alg:match} will return a sequence of all matchings possible with the passed arguments.
If implemented as described, it would compute this sequence eagerly and hence perform a lot of later possibly discarded computations.
In order for the implementation to be efficient, it is necessary to ensure that the sequence is computed lazily.
If available, we suggest using lazy sequences or co-routines provided by the host language.\footnote{See, for example, \url{https://docs.python.org/3/glossary.html\#term-generator}}

Next, we define the the \textsc{rule} and \textsc{compose} functions.
The functions construct a list of functions, which represent individual rule applications.
The \textsc{compose} function in \autoref{alg:instance:edsl} merely concatenates all subprograms.
\footnote{Note, that we define \textsc{compose} as a variadic function instead of a binary operator.}
The \textsc{rule} function constructs a singleton list with a function performing application of the rule if possible.

\begin{algorithm*}[t]
    \begin{algorithmic}[1]
        \Function{rule}{$\mathit{name}$, $\mathit{kept}$, $\mathit{removed}$, $\mathit{guard}$, $\mathit{body}$}
            \Function{rule\_program}{$i_a$, $c_a$,$\mathit{state}$}
                \State $(r_a, m_a) \leftarrow \mathtt{get\_active\_iterator}(\mathit{state})$
                    \algcomment{fetch iterator}

                \If{$m_a = \bot$}
                    \State $m_a \leftarrow \textsc{match}\left(\mathit{kept}\diamond\mathit{removed}, \mathit{guard}, i_a, c_a, \mathit{state}\right)$
                        \algcomment{init. iterator}
                \EndIf

                \If{$m_a \neq []$}
                    \State \Return $\mathtt{set\_active\_iterator}\left(\mathit{state}, (r_a+1, \bot)\right)$ 
                        \algcomment{iterator empty}
                \EndIf

                \State $(p_a, i_1, ..., i_n, v_1, ..., v_n) : m_a' \leftarrow m_a$
                    \algcomment{fetch one matching}
                \If{$\left|\mathit{kept}\right| \leq p_a$}
                        \algcomment{active value matched to $\mathit{removed}$?}
                    \State $\mathit{state} \leftarrow \mathtt{pop\_query}\left(\mathit{state}\right)$
                \Else
                    \State $\mathit{state} \leftarrow \mathtt{set\_active\_iterator}\left(\mathit{state}, (r_a, m_a')\right)$
                \EndIf

                \State $\mathit{state} \leftarrow \mathtt{remove}\left(\mathit{state}, i_{\left|\mathit{kept}+1\right|}, ..., i_n\right)$
                    \algcomment{remove values matched to $\mathit{removed}$}
                \State $b_1, ..., b_k \leftarrow \mathit{body}\left(v_1, ..., v_n\right)$
                \State $\mathit{state} \leftarrow \mathtt{push\_query}\left(\mathit{state}, b_1, ..., b_k\right)$
                    \algcomment{query new values}
                \State \Return $\mathit{state}$
                    \algcomment{return successor state}
            \EndFunction

            \State \Return $\left[\textsc{rule\_program}\right]$
        \EndFunction
        \State
        \Function{compose}{$p_1, ..., p_n$}
            \Return $p_1 \diamond ... \diamond p_n$
        \EndFunction
    \end{algorithmic}
    \caption{Concrete implementations for $\chrrule$ and $\chrcomp$}
    \label{alg:instance:edsl}
\end{algorithm*}
The outer functions gets all components of a FreeCHR rule passed, \IE, its $\mathit{name}$ (which is unused), $\mathit{kept}$ and $\mathit{removed}$ head, $\mathit{guard}$ and $\mathit{body}$.
It returns a single rule program.
The inner function gets the currently active value $(i_a, c_a)$ and the current state passed.
It first fetches the iterator of the currently active value and initializes it if necessary by calling the \textsc{match} function.
We then check if the current iterator is empty.
If this is the case, there are no (more) applicable matchings.
We hence increment the rule index of the iterator and return the updated state.
Otherwise, we read a matching from it.
We check if the active value was matched to a removed pattern.
If so, we drop the value from the query since it will be removed from the store and can no longer be subject to rule application.
Otherwise, we need to update the active iterator.
We then remove the removed values from the store and add the values computed by the body to the query.
The outer function then returns the inner function in a singleton list.


\begin{algorithm*}[t]
    \begin{algorithmic}[1]
        \Function{step}{$\left[p_1, ..., p_n\right]$, $\mathit{state}$}
            \If{$\mathtt{query}(state) = \left[\right]$}
                \Return $\mathit{state}$
                    \algcomment{final state}
            \EndIf

            \State $\left(d_a, c_a\right):\_ \leftarrow \mathtt{query}\left(\mathit{state}\right)$
            
            \If{$d_a = \bot$}
                \Return $\mathtt{activate}\left(\mathit{state}\right)$
                    \algcomment{activate value}
            \EndIf
        
            \State $(i_a, r, \_) \leftarrow d_a$

            \If{$\neg \mathtt{alive}\left(\mathit{state}, i_a\right) \vee r > n$}
                \Return $\mathtt{pop\_query}\left(\mathit{state}\right)$
                    \algcomment{drop active value}
            \EndIf

            \State \Return $p_{r}\left(i_a, c_a, \mathit{state}\right)$
                \algcomment{apply rule}
        \EndFunction
        \State 
        \Function{run}{$p, s$}
            \While{$\mathtt{query}(s) \not\equiv \left[\right]$}
                $s \leftarrow \textsc{step}\left(p, s\right)$
                    \label{alg:instance:run:apply}
            \EndWhile
            \State \Return $s$
                \label{alg:instance:run:return}
        \EndFunction
    \end{algorithmic}
    \caption{Driver functions}
    \label{alg:instance:driver}
\end{algorithm*}

Finally, we introduce the two driver functions \textsc{step} and \textsc{run} in \autoref{alg:instance:driver}.
The function \textsc{step} transforms a list of rule functions ($\tnat \setprod C \setprod \Omega_I C \rightarrow \flist \Omega_I C$) to an applicable function ($\Omega_I C \rightarrow \Omega_I C$) over states.
It accepts the program as a list of rule functions and the current state.
First, it checks if the query is empty.
In this case, a final state is reached and returned.
It then fetches the currently active value and its decoration from the query.
If it was not yet activated, it activates it.
Otherwise, it is checked if it is still alive and its accompanying rule index does not exceed the length of the program.
Otherwise it is dropped from the query.
Finally, the $r^\text{th}$ rule is applied to the current state.

The \textsc{run} function applies the program $p$ to the state until a final state is reached.

  \section{Evaluation}\label{sec:eval}
In this section, we want to evaluate the optimizations introduced in \autoref{sec:exec} and compare them to variants with disabled optimizations, as well as the CHR implementation provided by \emph{SWI-Prolog}, by performing benchmarks on several (Free)CHR programs.

\subsection{Experiment Setup}
To evaluate the effect of the presented optimizations we run benchmarks on example programs on our implementation of FreeCHR in \emph{Python} in three different configurations:
no optimizations (\textsc{freechr}), iterator based without manual indexing (\textsc{freechr-it}) and iterator based with manual indexing (\textsc{freechr-ix}).
To compare the runtime of FreeCHR with CHR, we also run benchmarks on variants of our programs with the CHR implementation provided by \emph{SWI-Prolog} (\textsc{chr-ref}).
For each program, we used randomly generated sets of 100 queries of the same problem and problem size.
For each set, we took the average recorded runtime, as well as the rate of queries completed within the time limit of one second.
For each program, we measured on sets with increasing problems sizes.
We generated problem instances for three different problems, where $n$ is the problem size.
\begin{itemize}
    \item Compute the greatest common divisor of a random number $x$ with $2 \leq x \leq 1000$ and $1000*n$. (\textsc{GCD}\ifthenelse{\boolean{preprint}}{, Appendix \ref{app:gcd}}{})
    \item Compute the shortest path between every two nodes in a graph with $2\times \lceil\sqrt{n}\rceil$ nodes and $n$ weighted edges. (\textsc{SHP}\ifthenelse{\boolean{preprint}}{, Appendix \ref{app:shp}}{})
    \item Compute the Levenshtein distance of two sequences of length $15$ with up to $\lceil 15*\frac{n}{100} \rceil$ random mutations. (\textsc{LEV}\ifthenelse{\boolean{preprint}}{, Appendix \ref{app:lev}}{})
\end{itemize}
\ifthenelse{\boolean{preprint}}{}{%
The implementations of the CHR\footnote{\url{https://gitlab.com/freechr/freechr-python-benchmarks/-/tree/submission-wlp2025/chr_programs?ref_type=tags}} as well as FreeCHR\footnote{\url{gitlab.com/freechr/freechr-py/-/tree/submission-wlp2025/src/examples?ref_type=tags}} programs can be found on \emph{GitLab} and in the appendix of the preprint \cite{rechenberger2025optimized-pre}.}
The benchmarks were executed on a Linux (Kernel version 6.14.6-2-MANJARO) machine with an AMD Ryzen 7 2700 with 8 cores with 3.200 GHz and 16 GB RAM. 
We used \emph{SWI-Prolog} version 9.3.9 with optimization flags \mintinline{text}{-O} and \mintinline{text}{--no-debug} and \emph{Python} version 3.13.3 and optimization flag \mintinline{text}{-OO}.

\subsection{Results}
The results of the benchmarks are shown in \autoref{tabel:results}.
The columns $t_r$, $t_0$, $t_{\mathit{it}}$ and $t_{\mathit{ix}}$ show the average runtime of a query in milliseconds.
The columns $c_r$, $c_0$, $c_{\mathit{it}}$ and $c_{\mathit{ix}}$ show the rate of queries completed within the time limit of one second.
The columns $\frac{t_0}{t_r}$, $\frac{t_{\mathit{it}}}{t_r}$ and $\frac{t_{\mathit{ix}}}{t_r}$ show the respective average runtime relative to $t_r$. 

\begin{table}[t]
    \centering
    \begin{tabularx}{\linewidth}{lc|*{2}{C}|*{3}{C}|*{3}{C}|*{3}{C}}
        \hline
        Prob. & Size & \multicolumn{2}{c}{\textsc{chr-ref}} & \multicolumn{3}{c}{\textsc{freechr}} & \multicolumn{3}{c}{\textsc{freechr-it}} & \multicolumn{3}{c}{\textsc{freechr-ix}} \\
                &      & $t_r$    & $c_r$ & $t_0$     & $\frac{t_0}{t_r}$ & $c_0$ & $t_{\mathit{it}}$     & $\frac{t_{\mathit{it}}}{t_r}$ & $c_{\mathit{it}}$ & $t_{\mathit{ix}}$     & $\frac{t_{\mathit{ix}}}{t_r}$ & $c_{\mathit{ix}}$ \\
        \hline\hline
        GCD & $10$ & $0.05$ & $1.0$ & $0.24$ &$4.69$ & $1.00$ & $0.26$ &$5.26$ & $1.00$ & $0.36$ &$7.17$ & $1.00$ \\
            & $100$ & $0.06$ & $1.0$ & $0.70$ &$12.3$ & $1.00$ & $0.52$ &$9.13$ & $1.00$ & $0.86$ &$15.3$ & $1.00$ \\
            & $1000$ & $0.12$ & $1.0$ & $0.99$ &$8.15$ & $1.00$ & $1.26$ &$10.3$ & $1.00$ & $1.61$ &$13.2$ & $1.00$ \\
            & $10000$ & $0.79$ & $1.0$ & $6.78$ &$8.53$ & $1.00$ & $9.28$ &$11.6$ & $1.00$ & $12.68$ &$15.9$ & $1.00$ \\
        \hline
        SHP & $10$ & $0.22$ & $1.0$ & $2.96$ &$13.2$ & $1.00$ & $2.01$ &$8.98$ & $1.00$ & $1.77$ &$7.91$ & $1.00$ \\
            & $20$ & $1.04$ & $1.0$ & $33.8$ &$32.4$ & $1.00$ & $17.3$ &$16.6$ & $1.00$ & $12.2$ &$11.7$ & $1.00$ \\
            & $40$ & $5.10$ & $1.0$ & $321$ &$63.1$ & $1.00$ & $125$ &$24.7$ & $1.00$ & $71.41$ &$14.01$ & $1.00$ \\
            & $80$ & $21.5$ & $1.0$ &  – &  –  & $0.00$ & $785$ &$36.4$ & $0.98$ & $388$ &$18.0$ & $1.00$ \\
        \hline
        LEV & $10$ & $36.0$ & $1.0$ & $402$ &$11.1$ & $0.54$ & $413$ &$11.4$ & $0.63$ & $393$ &$10.9$ & $0.64$ \\
            & $20$ & $31.1$ & $1.0$ & $344$ &$11.0$ & $0.59$ & $343$ &$11.0$ & $0.65$ & $357$ &$11.4$ & $0.69$ \\
            & $40$ & $32.4$ & $1.0$ & $327$ &$10.0$ & $0.55$ & $395$ &$12.1$ & $0.68$ & $391$ &$12.0$ & $0.71$ \\
            & $80$ & $32.6$ & $1.0$ & $324$ &$9.9$ & $0.55$ & $395$ &$12.1$ & $0.68$ & $392$ &$12.0$ & $0.71$ \\
        \hline
    \end{tabularx}
    \vskip.2cm
    \caption{Benchmark results}
    \label{tabel:results}
\end{table}

The first obvious result is that our FreeCHR implementation in Python is still slow (even with our optimizations) when compared to the SWI-Prolog implementation of CHR.
This is to some degree expected since the SWI-Prolog implementations is already very mature and FreeCHR is effectively an interpreted language, which comes with some penalties to performance.

Concerning our optimizations, we can generally see a positive effect on runtime.
The exception here is the GCD problem, where each optimization worsens runtime.
The reason for this is probably that since there are only two elements in the store at any given moment, finding a matching is already relatively easy.
The optimizations then only add overhead which slows down computation.
For the SHP problem, however, both optimizations increase execution speed significantly.
This can be seen on completion rate but more so on runtime.
The performance increase for the LEV problem becomes apparent when considering the completion rates.
Since the newly completed queries are probably relatively hard, \IE need more time to complete, the average runtime is increased.

Since our optimizations already have visible positive effects on performance (except for certain cases), and there is still much potential for optimizations, we consider those results very promising.

  \section{Conclusion and Future Work}\label{sec:conclusion}
In this paper, we introduced two optimizations, iterator based matching and manual indexing, which we designed to be applicable to FreeCHR.
We also provided an improved execution algorithm for FreeCHR and results of runtime benchmarks.
The results suggest the general effectiveness of the implemented optimizations concerning runtime.

Future work will be mostly concerned with further optimizations, as well as providing proofs of correctness, \WRT formally defined operational semantics \cite{rechenberger2025freechr,rechenberger2025refined}.

  \section*{Acknowledgements}
  We thank our anonymous reviewers for their very detailed and insightful reviews.
  They were of great help to improve the quality of the paper.

  \bibliographystyle{unsrturl}
  \bibliography{local/bibliography}

  \ifpreprint
  \appendix
  \chapter*{Appendix}

\section{Greatest Common Divisor (GCD)}\label{app:gcd}
\paragraph{SWI-Prolog}
\begin{minted}[autogobble]{prolog}
:- chr_constraint gcd(+int).

% Greatest common divisor in CHR(SWI-Prolog)
zero     @ gcd(0) <=> true.
subtract @ gcd(M) \ gcd(N) <=> 0 < M, M =< N | N1 is N - M, gcd(N1).
\end{minted}

\paragraph{Python}
\begin{minted}[autogobble]{python}
# Greatest common divisor in FreeCHR(Python)
gcd_solver = compose(
    rule("zero", [], [lambda x: x == 0], lambda _: True, lambda _: []),
    rule("subtract", [lambda n: 0 < n], [lambda m: 0 < m],
        (lambda n, m: n <= m),
        (lambda n, m: [m-n])
    )
)
\end{minted}

\section{Shortest Path (SHP)}\label{app:shp}
\paragraph{SWI-Prolog}
\begin{minted}[autogobble]{prolog}
:- chr_type list(T)
    ---> []
       ; [T | list(T)].

:- chr_constraint
    edge(+int, +int, +int),
    path(+int, +int, +int, +list(int)).

% Shortest Path algorithm in CHR(SWI-Prolog)
clear_duplicate_paths @
    path(X, Y, W, _) \ path(X, Y, W1, _) <=> W =< W1 | true.

primitive_paths @
    edge(X, Y, W) ==> path(X, Y, W, [X, Y]).

transitive_paths @
    edge(X, Y, W), path(Y, Z, Wp, Ps) ==>
        W1 is W+Wp,
        path(X, Z, W1, [X|Ps]).
\end{minted}

\paragraph{Python}
\begin{minted}[autogobble]{python}
# edge/3 constraints
@dataclass(frozen=True)
class Edge:
    source: str
    target: str
    weight: int

def is_edge(e):
    return isinstance(e, Edge) and not isinstance(e, Path)
\end{minted}

\begin{minted}[autogobble]{python}
# path/4 constraints
@dataclass(frozen=True)
class Path(Edge):
    nodes: list[str]

def is_path(p):
    return isinstance(p, Path) 
\end{minted}

\begin{minted}[autogobble]{python}
# Shortest-Path algorithm in FreeCHR(Python)
shortest_path_solver = compose(
    rule('clear duplicate paths',
        [is_path], [is_path],
        lambda e1, e2: e1.source == e2.source \
            and e1. target == e2.target \
            and e1.weight <= e2.weight,
        lambda *_: []
    ),
    rule('primitive paths',
        [is_edge], [],
        lambda _: True,
        lambda e: [Path(
            e.source, e.target,
            e.weight,
            [e.source, e.target]
        )]
    ),
    rule('transitive paths',
        [is_edge, is_path], [],
        lambda e, p: e.target == p.source,
        lambda e, p: [Path(
            e.source, p.target,
            e.weight + p.weight,
            [e.source] + p.nodes
        )]
    )
)
\end{minted}

\section{Levenshtein (LEV)}\label{app:lev}

\subsection*{Variable Assignments}

\paragraph{SWI-Prolog}
\begin{minted}[autogobble]{prolog}
:- op(700, xfx, (:=)).
:- chr_constraint :=(+,+).

% :=/2 solver in CHR(SWI-Prolog)
idem  @ X := V \ X := V <=> true.
trans @ Y := V \ X := Y <=> X := V.
fully_resolved @ 
    L := 1 + min(V1, V2, V3) <=> integer(V1), integer(V2), integer(V3) |
        V is 1 + min(V1, min(V2, V3)),
        L := V.
resolve_1 @ 
    L1 := V1 \ L := 1 + min(L1, L2, L3) <=> integer(V1) |
        L := 1 + min(V1, L2, L3).
resolve_2 @ 
    L2 := V2 \ L := 1 + min(L1, L2, L3) <=> integer(V2) |
        L := 1 + min(L1, V2, L3).
resolve_3 @ 
    L3 := V3 \ L := 1 + min(L1, L2, L3) <=> integer(V3) |
        L := 1 + min(L1, L2, V3).
\end{minted}

\paragraph{Python}
\begin{minted}[autogobble]{python}    
# :=/2 constraints
@dataclass
class Assignment(Indexed):
    lhs: str
    rhs: str | int | tuple[str | int, str | int, str | int]

    def index(self):
        return self.lhs

def is_assignment(a):
    return isinstance(a, Assignment)
\end{minted}

\begin{minted}[autogobble]{python}
# :=/2 solver in FreeCHR(Python)
assignment_solver = compose(
    rule('idem',
        [IndexedBy(1, lambda y: y.lhs, is_assignment)],
        [is_assignment],
        lambda x, y: x == y,
        lambda x, y: []
    ),
    rule('trans', 
        [IndexedBy(1, lambda y: y.rhs, is_assignment)],
        [lambda y: is_assignment(y) and isinstance(y.rhs, str)],
        lambda x, y: x.lhs == y.rhs,
        lambda x, y: [Assignment(y.lhs, x.rhs)]
    ),
    rule('fully resolved',
        [],
        [lambda x: is_assignment(x) \
            and isinstance(x.rhs, tuple) \
            and all(isinstance(v, int) for v in x.rhs)],
        lambda _: True,
        lambda x: [Assignment(x.lhs, 1+min(*x.rhs))]
    ),
    rule('resolve 0',
        [IndexedBy(1, lambda y: y.rhs[0],
            lambda x: is_assignment(x) and isinstance(x.rhs, int))],
        [lambda y: isinstance(y, Assignment) and isinstance(y.rhs, tuple)],
        lambda x, y: x.lhs == y.rhs[0],
        lambda x, y: [Assignment(y.lhs, (x.rhs, y.rhs[1], y.rhs[2]))]
    ),
    rule('resolve 1',
        [IndexedBy(1, lambda y: y.rhs[1],
            lambda x: is_assignment(x) and isinstance(x.rhs, int))],
        [lambda y: isinstance(y, Assignment) and isinstance(y.rhs, tuple)],
        lambda x, y: x.lhs == y.rhs[1],
        lambda x, y: [Assignment(y.lhs, (y.rhs[0], x.rhs, y.rhs[2]))]
    ),
    rule('resolve 2',
        [IndexedBy(1, lambda y: y.rhs[2],
            lambda x: is_assignment(x) and isinstance(x.rhs, int))],
        [lambda y: isinstance(y, Assignment) and isinstance(y.rhs, tuple)],
        lambda x, y: x.lhs == y.rhs[2],
        lambda x, y: [Assignment(y.lhs, (y.rhs[0], y.rhs[1], x.rhs))]
    ),
)
\end{minted}

\subsection{Levenshtein distance}
\paragraph{SWI-Prolog}
\begin{minted}[autogobble]{prolog}
:- chr_type list(T) ---> [] ; [T | list(T)].
:- chr_constraint ldist(+list(int), +list(int), +).

% Levenshtein Distance algorithm in CHR(SWI-Prolog)
memoization @ 
    ldist(Xs, Ys, L1) \ ldist(Xs, Ys, L2) <=> atom(L2) ; string(L2) |
        L2 := L1.
equation_1 @ 
    ldist(Xs, [], L) ==>
        length(Xs, _L),
        L := _L.
equation_2 @ 
    ldist([], Ys, S) ==> 
        length(Ys, L), 
        S := L.
equation_3 @ 
    ldist([X|Xs], [X|Ys], L) ==> 
        ldist(Xs, Ys, L).
equation_4 @ 
    ldist([X|Xs], [Y|Ys], L) ==> X \= Y |
        maplist(gensym(length), [L1, L2, L3]),
        L := 1 + min(L1, L2, L3),
        ldist(Xs, [Y|Ys], L1),
        ldist([X|Xs], Ys, L2),
        ldist(Xs, Ys, L3).
\end{minted}

\paragraph{Python}
\begin{minted}[autogobble]{python}
# ldist/2 constraints
@dataclass
class Levenshtein(Indexed):
    seq_a: list[int]
    seq_b: list[int]
    result_var: str

    def index(self):
        return (str(self.seq_a), str(self.seq_b))

def is_levenshtein(l):
    return isinstance(l, Levenshtein)
\end{minted}

\begin{minted}[autogobble]{python}
class SymbolGenerator:
    def __init__(self):
        self.next= 0    
        
    def gensym(self, prefix):
        sym = prefix + str(self.next)
        self.next += 1
        return sym

GEN = SymbolGenerator()


def equation_4_body(x):
    v1 = GEN.gensym('var')
    v2 = GEN.gensym('var')
    v3 = GEN.gensym('var')
    return [
        Assignment(x.result_var, (v1, v2, v3)),
        Levenshtein(x.seq_a[1:], x.seq_b, v1),
        Levenshtein(x.seq_a, x.seq_b[1:], v2),
        Levenshtein(x.seq_a[1:], x.seq_b[1:], v3)
    ]

# Levenshtein distance algorithm in FreeCHR(Python)
levenshtein_solver = compose(
    assignment_solver,
    rule('memoization',
        [IndexedBy(1, lambda y: y.index, is_levenshtein)],
        [is_levenshtein],
        lambda x, y: x.seq_a == y.seq_a and x.seq_b == y.seq_b,
        lambda x, y: [Assignment(y.result_var, x.result_var)]
    ),
    rule('equation_1',
        [lambda x: is_levenshtein(x) and not x.seq_b], [],
        lambda _: True,
        lambda x: [Assignment(x.result_var, len(x.seq_a))]
    ),
    rule('equation_2',
        [lambda x: is_levenshtein(x) and not x.seq_a], [],
        lambda _: True,
        lambda x: [Assignment(x.result_var, len(x.seq_b))]
    ),
    rule('equation_3',
        [lambda x: is_levenshtein(x) \
            and x.seq_a and x.seq_b and x.seq_a[0] == x.seq_b[0]], [],
        lambda _: True,
        lambda x: [Levenshtein(x.seq_a[1:], x.seq_b[1:], x.result_var)]
    ),
    rule('equation_4',
        [lambda x: is_levenshtein(x) \
            and x.seq_a and x.seq_b and x.seq_a[0] != x.seq_b[0]], [],
        lambda _: True,
        equation_4_body
    )
)
\end{minted}
  \fi
\end{document}